\documentclass[preprint,showpacs,showkeys,amsmath,amssymb,pra]{revtex4-1}
\usepackage{graphicx}
\usepackage{dcolumn}
\usepackage{bm}

\begin{document}

\title{Indivisible quantum evolution of a driven open spin-$S$ system}

\author{Xiang Hao }
\affiliation{School of Mathematics and Physics, Suzhou University of Science and Technology, Jiangsu 215009, China\\}
\affiliation{Department of Physics, Renmin University of China, Beijing 100872, China}

\author{Xuefen Xu}
\affiliation{School of Mathematics and Physics, Jiangsu Teachers University of Technology, Jiangsu 213001, China}
\author{Xiaoqun Wang}
\altaffiliation{Corresponding author} \email{xiaoqunwang@ruc.edu.cn}
\affiliation{Department of Physics, Renmin University of China, Beijing 100872, China}

\date{\today}

\begin{abstract}
Using a recently proposed measure for divisibility of a dynamical map, we study the non-Markovian character of a quantum evolution of a driven spin-$S$ system weakly coupled to a bosonic bath. The complete tomographic knowledge about the dynamics of the open system is obtained by the time-convolutionless master equation in the secular approximation. The derived equation can be applied to a wide range of spin-boson models with the Hermitian or non-Hermitian coupling operator in the system-environment interaction Hamiltonian. Besides the influence of the environmental spectral densities, the tunneling energy of the system Hamiltonian can affect the measure of quantum non-Markovianity. It is found that the non-Markovian feature of a dynamical map of a high-dimension spin system is noticeable in contrast to that of a low-dimension spin system.
\end{abstract}

\pacs{03.65.Yz, 03.65.Ta, 03.65.Ud, 42.50.Lc}

\keywords{Non-Markovianity, spin-s system, quantum evolution}

\maketitle

\section{\label{sec1}Introduction}

The realistic dynamics of an open quantum system is expected to deviate partly from the Markovian evolution which is idealized under the influence of memoryless environments~\cite{Breuer04,Maniscalco06,Breuer07,Wolf08,Piilo09,Rebentrost09,Vacchini11}. The coupling of a system to an environment inevitably leads to the loss of information and dissipation of energy~\cite{Breuer072,Weiss99,Kampen07}. In the Born-Markov approximation, the decoherence process shows a monotonic decrease with time. However, many kinds of condensed matter systems will undergo non-Markovian evolutions because of complicate interactions with their surroundings~\cite{Ishizaki09,Engel07,Goan11}. The non-Marokovian dynamics of an open quantum system can play an important role in a variety of modern applications such as quantum optics ~\cite{Breuer072,Liu11}, solid-state controllable technology~\cite{Lai06}, quantum chemistry~\cite{Plenio98}, and quantum information~\cite{Aharonov06}. Under the influence of environments with memory effects, the non-Markovian dynamics may preserve the existence of quantum entanglement for longer time~\cite{Yu06,An07}. Therefore, it is necessary to efficiently quantify the measure of non-Markovianity.

Recently, several approaches have been put forward~\cite{Breuer09,Rivas10,Hou11,Lu10}. The measure based on the distinction of quantum states is used to describe a temporary flow of information from the environment back to the system. The non-Markovian behavior will be remarkable if the trace distances of evolving states increase~\cite{Breuer09}. This measure emphasizes the physical features of the system-environment model. According to the composition law, another measure was proposed to quantify the deviation from the divisible map. An indivisible map shows the emergence of the non-Markovian dynamics. The mathematical description of this measure is simple due to no optimization procedure~\cite{Rivas10,Hou11}. Meanwhile, some operational measures were offered, including the measure by assessing the increase of entanglement between an open system and an isolated ancilla~\cite{Rivas10} and the measure by the quantum Fisher information flow~\cite{Lu10}.

Many theoretical models of efficient two-level systems coupled to their environments were investigated to quantify the measure of non-Markovianity of the evolutions~\cite{Porras08,Laine10,Apollaro11,Rebentrost11,Haikka11,Chruscinski11,Clos12}. As is known, high-dimension spin systems or effective $d$-level systems extensively consist in the nature~\cite{Yang12,Thomale12,Schlottmann12}. It is interesting to understand the complicate evolutions of these open systems. On one hand, we perform a reasonable unitary transformation to establish the second-order time-convolutionless master equation for a driven spin-$S$ system. With the consideration of the interaction Hamiltonian between the system and environment, the master equation is valid for both Hermitian and non-Hermitian system coupling operator cases. This is one point of the motivations of our present work. On the other hand, we utilize the measure for the divisibility to quantify the the non-Markovianity of the dynamical map. Through the strict theoretical analysis, we study in detail how the dimension of open system affects the degree of quantum non-Markovianity. This is the other point of the motivations of our investigations.

The article is organized as follows: In Sec.~\ref{sec2}, we briefly describe a model of a driven spin-$S$ system coupled to a bosonic bath. In the eigenrepresentation of the spin-$S$ Hamiltonian, the evolution of the open system is obtained perturbatively up to the second order in the weak interaction between the system and bath. The divisibility of the dynamical map is introduced and calculated to quantify the degree of the deviation from a Markovian process in Sec. ~\ref{sec3}. Finally, in Sec.~\ref{sec4}, we summarize our results and draw some conclusions.

\section{\label{sec2}Evolution equations of a driven spin-$S$ system coupled to thermal bath}

\subsection{\label{sec2:level21}Hamiltonian of system-environment model}

Our system-environment model consists of an arbitrary driven spin-$S$ system which is weakly interacting with a thermal bosonic environment. The general total Hamiltonian is composed of three parts,
\begin{equation}
\label{eq1}
H=H_{S}+H_{E}+H_{I}~,
\end{equation}
where $H_{S}=\omega_{s}\vec{n}\cdot \vec{\hat{S}}$ describes the Hamiltonian of a driven spin-$S$ system. The parameter $\omega_{s}$ denotes the transition frequency between any two neighboring energy states. The unit vector $\vec{n}=(\sin \alpha \cos \varphi,~\sin \alpha \sin \varphi,~\cos \alpha)$ is used to characterize the tunneling element of the Hamiltonian. These angles satisfy that $\alpha \in [0,\pi)$ and $\varphi \in [0,2\pi)$. The operator of spin-$S$ is expressed by three components of $\vec{\hat{S}}=(\hat{S}_{x},\hat{S}_{y},\hat{S}_{z})$. Using the raising and lowering operators $\hat{S}_{\pm}=\hat{S}_{x}\pm i \hat{S}_{y}$, the Hamiltonian of a driven system is also written as,
\begin{equation}
\label{eq2}
H_{S}=\frac 12\omega_{s}{\bf [}\sin \alpha (\hat{S}_{+}e^{-i\varphi}+\hat{S}_{-}e^{i\varphi})+2\cos \alpha \hat{S}_{z}{\bf ]}~.
\end{equation}
$\hat{S}_{\pm}\vert m\rangle=\sqrt {(S\pm m+1)(S\mp m)}\vert m\pm 1\rangle$ where $\vert m\rangle$ is the eigenvector of $\hat{S}_{z}$ with the eigenvalue $m$. Here, the tunneling energy is mainly determined by the angle parameter $\alpha$.

To conveniently investigate the dynamics of a driven spin system, we turn $H_{S}$ into the form in the eigenrepresentation spanned by the basis $\{ \vert \tilde{m}\rangle, \tilde{m}=\tilde{S},\ldots,-\tilde{S} \}$,
\begin{equation}
\label{eq3}
\tilde{H}_{S}=U^{\dag}H_{S}U~,
\end{equation}
where the unitary transformation is expressed as,
\begin{eqnarray}
\label{eq4}
U&=&\sum_{m=-S}^{S}\vert \psi_{m}\rangle \langle \tilde{m}\vert~,\nonumber \\
&=&\left(
\begin{array}{cccc}
a_{S}^{\text{(S)}}& a_{S}^{\text{(S-1)}} & \cdots & a_{S}^{\text{(-S)}}\\
a_{S-1}^{\text{(S)}}& a_{S-1}^{\text{(S-1)}} & \cdots & a_{S-1}^{\text{(-S)}}\\
\vdots & \vdots & \ddots & \vdots\\
a_{-S}^{\text{(S)}}& a_{-S}^{\text{(S-1)}} & \cdots & a_{-S}^{\text{(-S)}}
\end{array}\right).
\end{eqnarray}
Here the normalization state $\vert \psi_{m}\rangle=\sum_{k}a_{k}^{\text{(m)}}\vert k \rangle$ is the eigenvector of the Hamiltonian $H_{S}$.

In regard to weak couplings of an open system to an environment, we may reasonably consider the environment as a collection of harmonic oscillators. The Hamiltonian of the heat bath is written as,
\begin{equation}
\label{eq5}
H_{E}=\sum_{j}\omega_{j}b_j^{\dag}b_j~,
\end{equation}
where $b_j^{\dag}$ and $b_j$ are, respectively, the creation and annihilation operator of the $j$th harmonic oscillator with the frequency $\omega_j$.

The interaction Hamiltonian between the system and environment is expressed as,
\begin{equation}
\label{eq6}
H_{I}=\sum_j(g_j \hat{\Pi} b_j^{\dag}+g_j^{\ast}\hat{\Pi}^{\dag}b_j)~.
\end{equation}
The two cases of the system coupling operator $\hat{\Pi}$ are considered here. In one case, the operator is Hermitian, i.e., $\hat{\Pi}^{\dag}=\hat{\Pi}$. The other case of a non-Hermitian operator $\hat{\Pi}^{\dag}\neq \hat{\Pi}$ is also involved. The weak coupling strength $g_j$ can be characterized by the environmental spectral density $J(\omega)=\sum_j \vert g_j \vert^{2}\delta(\omega-\omega_j)$~\cite{Breuer072}.

\subsection{\label{sec2:level22} Quantum evolution equation}

On the assumption of the weak system-environment couplings, the time-convolutionless master equation for the second-order perturbation after the unitary transformation $U$ is written in the interaction picture to describe the dynamics of the open system,
\begin{equation}
\label{eq7}
\frac {d}{dt}\tilde{\rho}(t)=-\frac {1}{\hbar^2}\int_{0}^{t}dt_{1} \text{Tr}_{E} {\bf [} \tilde{H}_{I}(t),[\tilde{H}_{I}(t_1),\tilde{\rho}(t)\otimes \rho_{E}]  {\bf ]}~,
\end{equation}
where the density matrix of the open system $\tilde{\rho}(t)=\text{Tr}_{E}[\tilde{\rho}_{\text{tot}}(t)]$ is obtained by partially tracing over the freedom degrees of the environment. The initial total state is assumed to be factorized as $\tilde{\rho}_{\text{tot}}(0)=\tilde{\rho}(0)\otimes \rho_{E}$ where $\rho_{E}=\exp(-H_{E}/k_{B}T)/\text{Tr}[\exp(-H_{E}/k_{B}T)]$ denotes the thermal equilibrium state of the environment. Here the Planck constant $\hbar$ and Boltzmann constant $k_{B}$ are taken to be one.

The general interaction Hamiltonian in the interaction picture is expressed as,
\begin{eqnarray}
\label{eq8}
\tilde{H}_{I}(t)&=&U^{\dag}e^{iH_{0}t}H_{I}e^{-iH_{0}t}U \nonumber \\
                &=&\sum_{j}\sum_{l,m}c_{l,m}g_j^{\ast}e^{i[\omega_j-(l-m)\omega_s]t}\sigma_{l,m}b_j^{\dag}+\text{h.c.}~,
\end{eqnarray}
where the Hamiltonian $H_{0}$ represents $H_{0}=H_{S}+H_{E}$. The transition operator is defined as $\sigma_{l,m}=\vert \tilde{l}\rangle \langle \tilde{m} \vert$ and the notation $\text{h.c.}$ denotes the Hermitian conjugate of the former term in Eq.~(\ref{eq8}). The coefficients $c_{l,m}$ are calculated by $U^{\dag}\hat{\Pi} U=\sum_{l,m}c_{l,m}\sigma_{l,m}$.

In the field of condensed matter physics, many kinds of effective $d$-level quantum systems or some high-dimension spin systems have their coherence time $\tau_{S}$ which may be much smaller than the correlation time of the thermal environment $\tau_{E}$~\cite{Yang12}. In the condition of $\tau_{S} \ll \tau_{E}$, we may reasonably adopt the secular approximation by neglecting the high-frequency oscillating terms. This secular approximation is almost equivalent to the rotating wave approximation in quantum optics~\cite{Breuer072}. The necessary derivation of the quantum evolution equation is shown in the appendix. The general expression of the quantum master equation for the dynamics of the driven spin-$S$ system is presented as,
\begin{widetext}
\begin{eqnarray}
\label{eq9}
\frac {d}{dt}\tilde{\rho}(t)=& &\sum_{l}(\Gamma_{l,l}+\tilde{\Gamma}_{l,l})\vert c_{l,m} \vert^2\cdot(\sigma_{l,l}\tilde{\rho}\sigma_{l,l}-\frac 12\sigma_{l,l}\tilde{\rho}-\frac 12\tilde{\rho}\sigma_{l,l})\nonumber \\
&+&\sum_{l\neq k}(\Lambda_{l,l}+\tilde{\Lambda}_{l,l}^{\ast}) c_{k,k}c_{l,l}^{\ast}\cdot \sigma_{l,l}\tilde{\rho}\sigma_{k,k}+\text{h.c.} \nonumber \\
&+&\sum_{l\neq m}(\Gamma_{l,m}\vert c_{l,m} \vert^2+\tilde{\Gamma}_{m,l}\vert c_{m,l} \vert^2)\cdot(\sigma_{ml,}\tilde{\rho}\sigma_{l,m}-\frac 12\sigma_{l,l}\tilde{\rho}-\frac 12\tilde{\rho}\sigma_{l,l})~.
\end{eqnarray}
\end{widetext}
The decay rates in Eq.~(\ref{eq9}) are obtained as,
\begin{eqnarray*}
\Lambda_{l,m}(t)&=&\int_{0}^{t}dt_1 \int d\omega J(\omega)r e^{i[\omega-(l-m)\omega_s](t-t_1)} \\
\tilde{\Lambda}_{l,m}(t)&=&\int_{0}^{t}dt_1 \int d\omega J(\omega)(r+1) e^{i[\omega-(l-m)\omega_s](t-t_1)}~, \\
\end{eqnarray*}
where the other decay rates $\Gamma_{l,m}=2\text{Re}(\Lambda_{l,m})$ and $\tilde{\Gamma}_{l,m}=2\text{Re}(\tilde{\Lambda}_{l,m})$. The symbol function $\text{Re}(\cdot)$ represents the real part of a complex number. $r=1/(e^{\omega/T}-1)$ signifies the mean number of the thermal environment at a finite temperature.

For the simplest example of $S=\frac 12$, the coefficients for the Hermitian coupling operator $\hat{\Pi}=\hat{S}_{z}$ are calculated by $c_{\frac 12 ,\frac 12}=-c_{\frac {-1}{2},\frac {-1}{2}}=\frac 12\cos \alpha$ and $c_{\frac 12,\frac {-1}{2}}=c^{\ast}_{\frac {-1}{2},\frac {1}{2}}=\frac 12e^{i\varphi}\sin \alpha$. Then, the evolution equation just describes the dynamics of famous spin-boson models. In the other case of non-Hermitian operator $\hat{\Pi}=\hat{S}_{-}$, the coefficients are also obtained as $c_{\frac 12,\frac 12}=-c_{\frac {-1}{2},\frac {-1}{2}}=\frac 12e^{i\varphi}\sin \alpha$, $c_{\frac 12,\frac {-1}{2}}=-e^{2i\varphi}\cos^{2}\frac {\alpha}{2}$, and $c_{\frac {-1}{2},\frac 12}=-\sin^2 \frac {\alpha}{2}$. This quantum master equation can characterize the dynamics of the spin system coupled to the environment in the form of the dipole-dipole interaction. This result coincides with the quantum master equation in Ref.~\cite{Haikka11}.

It is also interesting to present the quantum master equation for the dynamics of a driven spin-$1$ system. The unitary transformation for the case of $S=1$ is given as,
\begin{eqnarray}
\label{eq10}
U&=&\left(
\begin{array}{ccc}
\cos^2\frac {\alpha}{2}e^{-i\varphi}~& -\frac {\sqrt{2}}{2}\sin \alpha e^{-i\varphi}~& -\sin^2\frac {\alpha}{2}e^{-i\varphi}\\
       \frac {\sqrt{2}}{2}\sin \alpha~& \cos \alpha ~& \frac {\sqrt{2}}{2}\sin \alpha\\
       \sin^2\frac {\alpha}{2}e^{i\varphi}~& \frac {\sqrt{2}}{2}\sin \alpha e^{i\varphi}~& -\cos^2\frac {\alpha}{2}e^{i\varphi}
\end{array}\right).
\end{eqnarray}
For the example of $\hat{\Pi}=\hat{S}_{z}$, the coefficients in Eq.~(\ref{eq9}) are written as $c_{l,l}=l\cdot \cos\alpha,~c_{l,0}=c_{0,l}=-l\cdot\frac {\sqrt{2}}{2}\sin \alpha,(l=1,-1)$, and $c_{1,-1}=c_{-1,1}=c_{0,0}=0$. In this case, the expression of the quantum master equation is simplified as,
\begin{widetext}
\begin{eqnarray}
\label{eq11}
\frac {d}{dt}\tilde{\rho}(t)=& &\gamma_{0}\cdot[(\sigma_{1,1}-\sigma_{-1,-1})\tilde{\rho}(\sigma_{1,1}-\sigma_{-1,-1})-\frac 12(\sigma_{1,1}-\sigma_{-1,-1})\tilde{\rho}-\frac 12\tilde{\rho}(\sigma_{1,1}-\sigma_{-1,-1})]\nonumber \\
&+&\gamma_{+}\cdot[\sigma_{1,0}\tilde{\rho}\sigma_{0,1}+\sigma_{0,-1}\tilde{\rho}\sigma_{-1,0}-\frac 12\sigma_{0,0}\tilde{\rho}-\frac 12 \tilde{\rho}\sigma_{0,0}-\frac 12\sigma_{-1,-1}\tilde{\rho}-\frac 12 \tilde{\rho}\sigma_{-1,-1}] \nonumber \\
&+&\gamma_{-}\cdot[\sigma_{0,1}\tilde{\rho}\sigma_{1,0}+\sigma_{-1,0}\tilde{\rho}\sigma_{0,-1}-\frac 12\sigma_{0,0}\tilde{\rho}-\frac 12 \tilde{\rho}\sigma_{0,0}-\frac 12\sigma_{1,1}\tilde{\rho}-\frac 12 \tilde{\rho}\sigma_{1,1}] ~.
\end{eqnarray}
\end{widetext}
The decay rates in Eq.~(\ref{eq11}) are written as $\gamma_{0}=\vert c_{1,1}\vert^2(\Gamma_{1,1}+\tilde{\Gamma}_{1,1})$, $\gamma_{+}=\vert c_{0,1}\vert^2(\Gamma_{0,1}+\tilde{\Gamma}_{1,0})$, and $\gamma_{-}=\vert c_{1,0}\vert^2(\Gamma_{1,0}+\tilde{\Gamma}_{0,1})$.

\section{\label{sec3}Non-Markovianity measured by divisibility}

We will quantify the quantum non-Markovianity based on the divisibility of a dynamical map introduced by Rivas, Huelga, and Plenio~(RHP)~\cite{Rivas10}. According to the composition law, a trace-preserving completely positive map $\Theta(t,0)$ is divisible if it can be expressed as,
\begin{equation}
\label{eq12}
\Theta(t+t_1,0)=\Theta(t+t_1,t)\Theta(t,0)~,
\end{equation}
and $\Theta(t+t_1,t)$ is completely positive for any $t~,~t_1>0$. The divisible map $\Theta(t+t_1,0)$ is Markovian exactly. It is shown that the quantity satisfies,
\begin{equation}
\label{eq13}
g(t)=\lim_{t_1\rightarrow 0^{+}}\frac {\vert \vert\mathbb{I}_d\otimes\Theta(t+t_1,t) P_{d}^{+}\vert \vert_1-1}{t_1}\geq 0~,
\end{equation}
where $\mathbb{I}_d$ is the identity matrix in the $d$-dimension Hilbert space $\mathcal{H}^{d}$ and the density matrix $P_{d}^{+}$ signifies the maximally entangled state in the product Hilbert space $\mathcal{H}^{d\otimes d}$. The function $\vert \vert \text{A} \vert \vert_1$ represents the norm of the matrix $\text{A}$. The value of $g(t)$ is strictly positive if and only if the dynamical map $\Theta$ is indivisible. Therefore, the condition of $g(t)>0$ determines the emergence of the non-Markovian dynamics. The quantity $g(t)$ gives rise to a measure for quantum non-Markovianity defined as,
\begin{equation}
\label{eq14}
N_{RHP}(\Theta)=\frac {\mathcal{I}}{\mathcal{I}+1},~~\mathcal{I}=\int_{0}^{\infty}dt g(t)~.
\end{equation}
For the dynamical map of an open spin-$S$ system, the maximally entangled state is written as $P_{d}^{+}=\frac 1d\sum_{j,n=-S}^{S}\vert \tilde{j}\rangle\langle \tilde{n} \vert \otimes \vert \tilde{j}\rangle\langle \tilde{n} \vert$ where $d=2S+1$. When the dynamics of the open system is characterized by the quantum master equation $\partial_{t}\tilde{\rho}=\mathcal{L}_{t}\tilde{\rho}$, the key quantity $g(t)$ is expressed as,
\begin{equation}
\label{eq15}
g(t)=\lim_{t_1\rightarrow 0^{+}}\frac {\vert \vert\varepsilon(t,t_1)\vert \vert_1-1}{t_1}~,
\end{equation}
where the matrix $\varepsilon(t,t_1)=[(\mathbb{I}_{d}+t_1 \mathcal{L}_t)\otimes \mathbb{I}_d] P_{d}^{+}$~\cite{Rivas10}.

For the case of $S=1$, applying the dynamical map like Eq.~(\ref{eq11}), we can obtain the matrix $\varepsilon(t,t_1)$ as,
\begin{widetext}
\begin{eqnarray}
\label{eq16}
\varepsilon&=&\frac 13\left(
\begin{array}{ccccccccc}
1-t_1\gamma_{-}& 0 & 0 & 0~ & 1-t_1y~& 0 & 0& 0~&1-t_1v\\
0& t_1\gamma_+ & 0 & 0~ & 0~& 0 & 0& 0~&0\\
0& 0 & 0 & 0~ & 0~& 0 & 0& 0~&0\\
0& 0 & 0 & t_1\gamma_-~ & 0~& 0 & 0& 0~&0\\
1-t_1y& 0 & 0 & 0~ & 1-t_1(\gamma_++\gamma_-)~& 0 & 0& 0~&1-t_1w\\
0& 0 & 0 & 0~ & 0~& t_1\gamma_+ & 0& 0~&0\\
0& 0 & 0 & 0~ & 0~& 0 & 0& 0~&0\\
0& 0 & 0 & 0~ & 0~& 0 & 0& t_1\gamma_-~&0\\
1-t_1v& 0 & 0 & 0~ & 1-t_1w~& 0 & 0& 0~& 1-t_1\gamma_{+}
\end{array}\right),
\end{eqnarray}
\end{widetext}
where the elements are defined as $y=\frac{\gamma_0}{2}+\frac{\gamma_+}{2}+\gamma_-$, $v=\frac{\gamma_+}{2}+\frac{\gamma_-}{2}+2\gamma_0$, and $w=\frac{\gamma_0}{2}+\frac{\gamma_-}{2}+\gamma_+$. The key quantity $g(t)$ is calculated as,
\begin{equation}
\label{eq17}
g(t)=-\frac {4}{3}\sum_{j=0,\pm}\Phi(\gamma_j)~,
\end{equation}
where the value of the function is given by $\Phi(x)=x,~(x<0)$ and $\Phi(x)=0,~(x\geq 0)$.

To analytically study the effects of the dimension of the open system on the non-Markovianity, we can offer the general expression of the matrix $\varepsilon(t,t_1)$ as,
\begin{widetext}
\begin{eqnarray}
\label{eq18}
\varepsilon(t,t_1)=&&P_{d}^{+}+\frac {t_1}{d}\{\sum_{j,n,l}(\Gamma_{l,l}+\tilde{\Gamma}_{l,l}) \vert c_{l,l}\vert^2(\delta_{lj}\delta_{nl}\sigma_{l,l}-\frac 12\delta_{lj}\sigma_{ln}-\frac 12\delta_{nl}\sigma_{j,l})\nonumber \\
&+&\sum_{j,n,l\neq k}(\Lambda_{l,l}+\tilde{\Lambda}_{l,l}^{\ast})c_{l,l}^{\ast}c_{k,k}\delta_{lj}\delta_{nk}\sigma_{l,k}+\text{h.c.}\nonumber \\
&+&\sum_{j,n,l\neq m} \vert c_{l,m}\vert^2\Gamma_{l,m}(\delta_{lj}\delta_{nl}\sigma_{m,m}-\frac 12\delta_{lj}\sigma_{ln}-\frac 12\delta_{nl}\sigma_{j,l})\nonumber \\
&+&\sum_{j,n,l\neq m} \vert c_{l,m}\vert^2\tilde{\Gamma}_{l,m}(\delta_{mj}\delta_{nm}\sigma_{l,l}-\frac 12\delta_{nj}\sigma_{mn}-\frac 12\delta_{nm}\sigma_{j,m})\}\otimes \vert \tilde{j}\rangle \langle \tilde{n} \vert~,
\end{eqnarray}
\end{widetext}
where $\delta_{lj}=\delta_{jl}$ is a delta function. For the case of Hermitian coupling operator $\hat{\Pi}=\hat{S}_{z}$, the coefficients in Eq.~(\ref{eq18}) are calculated as $c_{l,l}=l\cdot \cos \alpha,~l=-S,\cdots,~S$. With respect to the special case of $\alpha \approx 0$, the decay rates $\Gamma_{l,l}=\Gamma_{0}$ and $\tilde{\Gamma}_{l,l}=\tilde{\Gamma}_{0}$ dominate in the dynamics of the open system because the off-diagonal elements $\vert c_{l,m}\vert^2\approx0,~(l\neq m)$ . In this condition, we can analytically obtain the quantity $g(t)$ as,
\begin{equation}
\label{eq19}
g(t)=-\frac {f(d)}{d}\cos^2 \alpha [\Phi(\Gamma_{0})+\Phi(\tilde{\Gamma}_{0})]~,
\end{equation}
where the quantity $g(t)$ is closely related to the dimension of the system and the function $f(d)$ satisfies $f(d+3)=3[f(d+2)-f(d+1)]+f(d)+1$ where $f(2)=1,~f(3)=4,~f(4)=10$. It is straightly seen that the degree of non-Markovianity can be increased with the dimension of spin system.

\section{\label{sec4}Conclusions}

We derive in detail the second-order time-convolutionless master equation for the dynamics of a driven spin-$S$ system weakly coupled to the thermal bosonic environment. The quantum evolution equation expressed in the eigenrepresentation of the Hamiltonian $H_{S}$ is valid for both Hermitian and non-Hermitian coupling operator in the interaction Hamiltonian $H_{I}$. We also present the quantum master equation for the case of $S=1$ in the secular approximation. The quantum evolution equation can be generally applied to the dynamics of any effective $d$-level quantum system. To conveniently characterize the degree of the deviation from Markovian dynamics, we analytically study the indivisibility of the dynamical map of a driven spin-$S$ system. The case of $S=1$ is also considered. The indivisibility of the map is mainly determined by the negative values of the decay rates and also related to the tunneling energy. Trough the theoretic analysis, we obtain the interesting result that the non-Markovian behavior of high-dimension systems may be remarkable compared to that of low-dimension systems.

\begin{acknowledgments}
The work is supported in part by the National Natural Science Foundation of China Grandts No. 10904104 and No. 11174114 and No. 11174363.
\end{acknowledgments}

\appendix

\section{Derivation of quantum master equation~
(\ref{eq9})}

In this appendix, we derive in detail the second-order two-convolutionless master equations~(\ref{eq7}) and (\ref{eq9}). Firstly, the interaction Hamiltonian $\tilde{H}_{I}(t)$ can be rewritten as,
\begin{equation}
\label{eq:app1}
\tilde{H}_{I}(t)=\hat{B}^{\dag}(t)\otimes \hat{C}(t)+\hat{B}(t)\otimes \hat{C}^{\dag}(t)~,
\end{equation}
where the operators are defined as $\hat{B}=\sum_j g_j e^{-i\omega_j t} b_j$ and $\hat{C}=\sum_{l,m}c_{l,m}e^{-i(l-m)\omega_s t}\sigma_{l,m}$. The dynamics of the total state of the system-environment model is calculated as,
\begin{equation}
\label{eq:app2}
\frac {d}{dt}\tilde{\rho}_{\text{tot}}=-\frac {i}{\hbar}[\tilde{H}_{I}(t),\tilde{\rho}_{\text{tot}}]-\frac {1}{\hbar^2}\int_{0}^{t}dt_{1} {\bf [} \tilde{H}_{I}(t),[\tilde{H}_{I}(t_1),\tilde{\rho}_{\text{tot}}(t_1)]  {\bf ]}~.
\end{equation}
On the assumption of the factorized initial total state $\tilde{\rho}(0)\otimes \rho_{E}$, we may safely make the approximation of $\text{Tr}_{E}[\tilde{H}_{I}(t)\rho_{E}]=0$ in order to eliminate the first term in Eq.~(\ref{eq:app2})~\cite{Goan11}. When the Born approximation is adopted, we can replace $\tilde{\rho}_{\text{tot}}(t_1)$ with $\tilde{\rho}(t)\otimes \rho_{E}$ to obtain Eq.~(\ref{eq7}) after performing the partial trace over the freedom degrees of the environment.

Then, the second-order time-convolutionless master equation can be expanded as,
\begin{widetext}
\begin{eqnarray}
\label{eq:app3}
\frac {d}{dt}\tilde{\rho}(t)=&& \int_{0}^{t}dt_1\int d\omega J(\omega)r\cdot e^{i\omega(t-t_1)}{\bf [} \hat{C}^{\dag}(t_1)\tilde{\rho}\hat{C}(t)-\hat{C}(t)\hat{C}^{\dag}(t_1)\tilde{\rho} {\bf ]} \nonumber \\
&+&\int_{0}^{t}dt_1\int d\omega J(\omega)(r+1)\cdot e^{i\omega(t-t_1)}{\bf [} \hat{C}(t)\tilde{\rho}\hat{C}^{\dag}(t_1)-\tilde{\rho}\hat{C}^{\dag}(t_1)\hat{C}(t) {\bf ]}+\text{h.c.}
\end{eqnarray}
\end{widetext}
The term of $\hat{C}^{\dag}(t_1)\tilde{\rho}\hat{C}(t)$ in Eq.~(\ref{eq:app3}) is obtained as,
\begin{equation*}
\hat{C}^{\dag}(t_1)\tilde{\rho}\hat{C}(t)=\sum_{l,m}\sum_{k,j}c_{l,m}^{\ast}c_{k,j}e^{i[(l-m)t_1+(j-k)t]\omega_s }\sigma_{m,l}\tilde{\rho}\sigma_{k,j}~.
\end{equation*}
When the secular approximation of $k=l \neq j=m$ and $k=j \neq l=m$ is made, we can simplify the above equation as,
\begin{widetext}
\begin{equation}
\label{eq:app4}
\hat{C}^{\dag}(t_1)\tilde{\rho}\hat{C}(t)=\sum_{l,k}c_{l,l}^{\ast}c_{k,k}\sigma_{l,l}\tilde{\rho}\sigma_{k,k}+\sum_{l\neq m}\vert c_{l,m} \vert^2e^{-i(l-m)\omega_s(t-t_1)}\sigma_{m,l}\tilde{\rho}\sigma_{l,m}~.
\end{equation}
\end{widetext}
Similarly, other terms in Eq.~(\ref{eq:app3}) are written as,
\begin{widetext}
\begin{eqnarray}
\label{eq:app5}
\hat{C}(t)\tilde{\rho}\hat{C}^{\dag}(t_1)&=&\sum_{l,k}c_{l,l}^{\ast}c_{k,k}\sigma_{k,k}\tilde{\rho}\sigma_{l,l}+\sum_{l\neq m}\vert c_{l,m} \vert^2e^{-i(l-m)\omega_s(t-t_1)}\sigma_{l,m}\tilde{\rho}\sigma_{m,l}\nonumber \\
\hat{C}(t)\hat{C}^{\dag}(t_1)\tilde{\rho}&=&\sum_{l,k}\delta_{kl}c_{l,l}^{\ast}c_{k,k}\sigma_{k,k}\tilde{\rho}+\sum_{l\neq m}\vert c_{l,m} \vert^2e^{-i(l-m)\omega_s(t-t_1)}\sigma_{l,l}\tilde{\rho}\nonumber \\
\tilde{\rho}\hat{C}^{\dag}(t_1)\hat{C}(t)&=&\sum_{l,k}\delta_{lk}c_{l,l}^{\ast}c_{k,k}\tilde{\rho}\sigma_{l,l}+\sum_{l\neq m}\vert c_{l,m} \vert^2e^{-i(l-m)\omega_s(t-t_1)}\tilde{\rho}\sigma_{m,m}~,
\end{eqnarray}
\end{widetext}
where $\delta_{kl}=\delta_{lk}$ is a delta function. By substituting Eqs.~(\ref{eq:app4}) and (\ref{eq:app5}) into Eq.~(\ref{eq:app3}), we can obtain the quantum master equation of an open spin-$S$ system like Eq.(\ref{eq9}).

\bibliography{manuscript}

\end{document}